\documentclass[12pt,preprint]{aastex}

\usepackage{psfig}
\usepackage{amsmath}

\newcommand{\ltsima}{$\buildrel<\over\sim$}
\newcommand{\lapprox}{\lower.5ex\hbox{\ltsima}}

\shorttitle{A Strongly Lensed Galaxy at z$\sim$7}
\shortauthors{Kneib et al.}

\received{2004 January 7}
\begin{document}

\title{A Probable z$\sim$7 Galaxy Strongly Lensed by the Rich Cluster
Abell\,2218: Exploring the Dark Ages}

\author{Jean-Paul Kneib\altaffilmark{1,2},
Richard S. Ellis\altaffilmark{2}, Michael R. Santos
\altaffilmark{2,3} Johan Richard\altaffilmark{1,2}}
\altaffiltext{1}{Observatoire Midi-Pyr\'en\'ees, UMR5572,
     14 Avenue Edouard Belin, 31000 Toulouse, France}
\altaffiltext{2}{Caltech, Astronomy, 105-24, Pasadena, CA 91125, USA}
\altaffiltext{3}{Institute of Astronomy, Madingley Road, Cambridge, CB3 0HA, UK}

\footnotetext{\footnotesize
Using data obtained with the Hubble Space Telescope operated by
AURA for NASA and the W.M. Keck Observatory on Mauna Kea, Hawaii.
The W.M. Keck Observatory is operated as a scientific partnership
among the California Institute of Technology, the University of
California and NASA and was made possible by the generous
financial support of the W.M. Keck Foundation.}


\begin{abstract}

We discuss the observational properties of a remarkably faint
triply-imaged galaxy revealed in a deep $z'$-band Advanced Camera
for Surveys observation of the lensing cluster Abell 2218
($z=$0.175). A well-constrained mass model for the cluster, which
incorporates the outcome of recent Keck spectroscopic campaigns,
suggests that the triple system arises via a high redshift ($z>6$)
source viewed at high magnification ($\simeq\times$25). Optical
and infrared photometry from Hubble Space Telescope and the Keck
Observatory confirms the lensing hypothesis and suggests a
significant discontinuity occurs in the spectral energy
distribution within the wavelength interval 9250--9850\AA. If this
break is associated with Gunn-Peterson absorption from neutral
hydrogen, a redshift of 6.6$\,<z<$\,7.1 is inferred. Deep Keck
spectroscopy conducted using both optical and infrared
spectrographs fails to reveal any prominent emission lines in this
region. However, an infrared stellar continuum is detected whose
decline below 9800\AA\ suggests a spectroscopic redshift towards
the upper end of the range constrained photometrically, i.e.
$z\simeq$7. Regardless of the precise redshift, the source is
remarkably compact ($\lapprox 1\,h_{70}^{-1}$kpc) and faint
($z_{F850LP}=$28.0) yet is undergoing vigorous star formation at a
rate $\simeq$2.6~$M_{\odot}$yr$^{-1}$. An intriguing property is
the steep slope of the ultraviolet continuum implied by the
photometry which may suggest that the source is representative of
an early population of galaxies responsible for cosmic reionization.
Independent verification of these results is highly desirable but
our attempts highlight the difficulty of studying such sources
with present facilities and the challenges faced in pushing back
the frontiers of the observable universe beyond $z\sim$6.5.

\end{abstract}


\section{Introduction}

Reionization was a landmark event which imprinted a signature over
the scale of the entire universe. After decades of lower limits on
the redshift at which it occurred, recent observations of QSOs
discovered by the Sloan Digital Sky Survey (Becker et al.\ 2001;
Djorgovski et al.\ 2001, Fan et al.\ 2002) suggest that
reionization was just finishing at $z\sim6-6.5$. The discovery of
$z\simeq6.5$ galaxies with strong Lyman~$\alpha$ emission (Hu et
al.\ 2002, Kodaira et al.\ 2003) is illustrative of possible
sources which may be responsible. Analysis of recent temperature
and polarization fluctuation data from the \textit{WMAP} satellite
suggests an optical depth for Thompson scattering of
$\tau=0.17\pm$0.04, implying that reionization began at higher
redshift, perhaps as early as $z\sim 15-20$ (Kogut et al.\ 2003,
Spergel et al.\ 2003).

The discovery of star-forming galaxies at $z\simeq6.5$ is an
important step toward understanding the nature of the sources
responsible for the end of cosmic reionization. However, to
explore the earlier stages implied by the \textit{WMAP} results,
it is necessary to push the search for star forming systems to
higher redshifts. With current facilities this is technically very
challenging. It is perhaps salutary to note that the current
redshift frontier ($z=6.5$, corresponding to an observed
Lyman~$\alpha$ wavelength of 9200\AA ) is coincident with the
wavelength at which optical CCD detectors fall significantly in
their quantum efficiency. Ground-based infrared spectroscopy,
necessary for exploring sources at higher redshift, is especially
difficult at faint limits.

Color-based searches for $z>$6 sources with the Advanced Camera
for Surveys ({\em ACS}) on-board the Hubble Space Telescope ({\em
HST}) is now recognized as a valuable way of locating $z>6$
sources. Promising results have been obtained by utilizing the
long wavelength F850LP ($z'$-band) filter in conjunction with deep
infrared imaging on {\em HST} or with large ground-based
telescopes (Bouwens et al.\ 2003, Yan et al.\ 2003, Stanway et
al.\ 2003, Dickinson et al.\ 2003). In view of significant
contamination of red `drop-out' samples by cool stars (e.g.
Stanway et al.\ 2003), the primary challenge lies in
spectroscopically verifying faint candidates (e.g. Dickinson et
al.\ 2000). Neither optical nor infrared spectrographs on the
current generation of ground-based telescopes may have the
sensitivity to give convincing results unless strong emission
lines are present. Although most of the distant sources found
beyond $z\simeq$5 have been identified via strong Lyman~$\alpha$
emission (Stern \& Spinrad 1999, Spinrad 2003), some sources
should have weak or no Lyman~$\alpha$ emission (e.g. Spinrad et
al.\ 1998). Indeed, a significant fraction ($\sim 75$\%) of the
most intensely star-forming galaxies located by color selection
techniques at $z\simeq3$, reveal Lyman $\alpha$ only in absorption
(Shapley et al.\ 2003).

Gravitational magnification by foreground clusters of galaxies,
whose mass distributions are tightly constrained by arcs and
multiple images of known redshift, has already provided new
information on the abundance of high redshift objects (Kneib et al
1996, Santos et al.\ 2003). Particularly high magnifications
($\simeq\times$10-50) are expected in the {\it critical regions}
which can be located precisely in well-understood clusters for
sources occupying specific redshift ranges (Ellis et al.\ 2001).
Although the volumes probed in this way are far smaller than those
addressed in panoramic narrow band surveys (Hu et al.\ 1998,
Malhotra et al.\ 2001, Hu et al.\ 2003) or the color-based surveys
cited earlier, if the surface density of background sources is
sufficient, lensing may provide the necessary boost for securing
the first glimpse of young cosmic sources beyond $z\simeq$6.5
(Santos et al.\ 2003).

In the course of studying the detailed rest-frame properties of
the image pair of a lensed $z=5.576$ galaxy in the cluster
Abell~2218 (Ellis et al 2001), we have discovered a new faint pair
of images in a deep $z'$-band {\em ACS} observation of this
cluster. Based on the geometrical configuration of this pair and
its photometric properties, we concluded that the images most
likely arise via strong magnification of a distant $z>$6 source.
The implied high redshift of this source led us to explore its
properties in more detail.

A plan of the paper follows. We present the photometric
observations in Section 2. Section 3 discusses redshift
constraints determined independently from the lensing model of
Abell 2218 and the spectral energy distribution based on
broad-band photometry. Section 4 summarizes our attempts to detect
Lyman~$\alpha$ emission spectroscopically and discusses the
implications of a continuum discontinuity seen in the infrared
spectrum. We discuss the source properties and implications
further in Section 5. Throughout we assume a cosmological model
with $\Omega_{\rm M}=0.3$, $\Omega_\Lambda=0.7$ and H$_0$=70
km~s$^{-1}$~Mpc$^{-1}$.

\section{Photometric Observations}

The source in question was originally discovered as a pair of
images ($a$ and $b$ on Figure~1) with reflection symmetry and a
separation of 7\arcsec\, in a deep (5-orbit, 11.31 ksec) {\em HST}/{\em
ACS}-F850LP ($z'$-band) observation of the Abell 2218 cluster of
galaxies. The observation was conducted as part of GO program 9452
(PI: Ellis) to characterize the stellar continuum in the lensed
pair at $z$=5.576 discussed earlier by Ellis et al.\ (2001).

The new {\em ACS} images were reduced using standard {\sc
IRAF}\footnote{{\sc IRAF} is distributed by the National Optical
Astronomy Observatories, which are operated by the Association of
Universities for Research in Astronomy, Inc., under cooperative
agreement with the National Science Foundation.} and {\sc STSDAS}
routines. The source was found by blinking the reduced F850LP
image with two archival images taken at shorter wavelength with
the F814W and F606W filters of the Wide Field Planetary Camera 2
({\em WFPC2}) (SM-3a ERO program 8500, PI: Fruchter; exposure time
F606W: 10 ksec, F814W: 12 ksec). A prominent discontinuity in the
brightness of the pair can be seen viewing, in sequence, the
F850LP, F814W and F606W frames (Figure~1). Together with the
geometrical arrangement and symmetry of the pair in the context of
our mass model for Abell 2218, this suggested the source is at
high redshift and gravitationally magnified by the cluster (see
$\S$3).

The two images have similar appearance in the {\em HST}/NICMOS
F160W image. This $H$-band data was also obtained under the GO
program 9452 and was acquired in a 4-orbit CVZ observation
totalizing 22.2~ksec of exposure time.  Similarly, we also
identified the lensed pair in a deep Keck NIRC $J$-band image
(5.64 ksec exposure) taken in 0.65\arcsec seeing on July 22 and
23, 2002 (Blain and Reddy private communication).

A clear mirror symmetry is seen in the {\em HST} data; both images
contain a bright core, a second fainter knot and extended emission
of lower surface brightness. The orientation of the pair also
closely matches the predicted shear direction (Figure 1). The
lensing hypothesis is further verified by comparing the colors of
the two images $a$ and $b$ using the available photometric data
summarized in Table 1\footnote{All photometric quantities are
based on the Vega system}. Colors were computed using a fixed
elliptical aperture, and are identical for both images within the
uncertainties. (For the $z_{850LP}-J$ color, the {\em ACS}
$z'$-band image was convolved by a gaussian to match the seeing of
the Keck NIRC J image).

\clearpage
\begin{figure*}
\centerline{\psfig{file=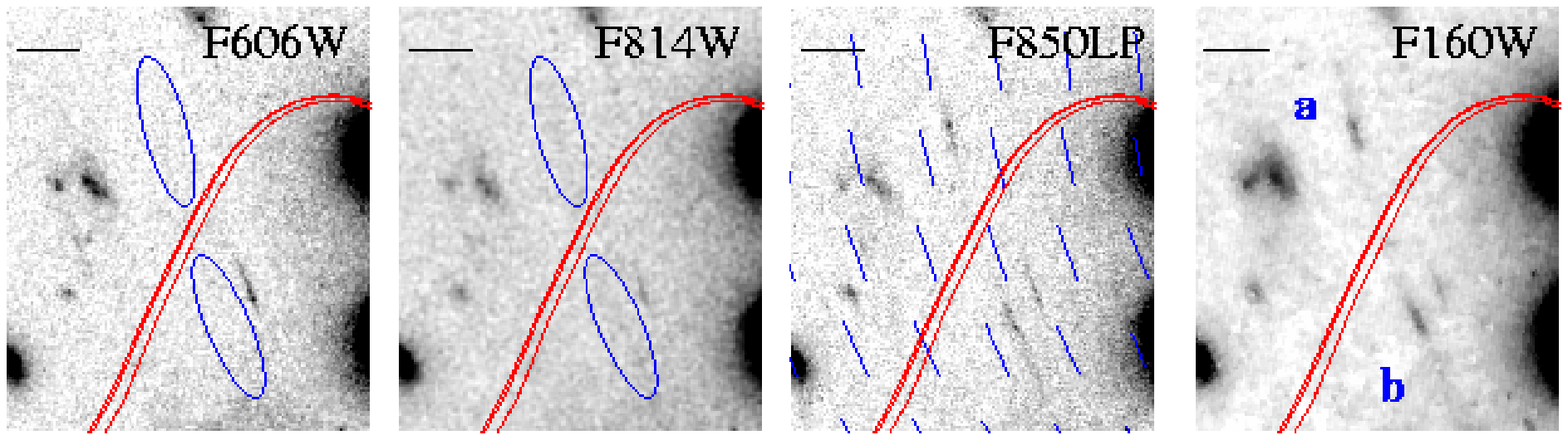,width=18cm}}

\noindent{\bf Figure 1:}\quad {\em WFPC2}-F606W, {\em
WFPC2}-F814W, {\em ACS}-F850LP and NICMOS-F160W images of Abell
2218 of the new faint pair in the lensing cluster Abell 2218
($z$=0.175). The signals redward of the {\em WFPC2}-F814W
observation suggests a marked break occurs in the continuum signal
at around 9600\AA\ . Red lines correspond to the predicted
location of the critical lines at $z_s$=5,6.5 and 7 (from bottom
to top, the latter two being almost coincident). The scale bar at
the top left of each image represents 2\arcsec. The predicted
shear direction (thin blue lines) closely matches the orientation of
the lensed images.
\end{figure*}

\clearpage
\begin{figure*}
\centerline{see jpeg file}
 \noindent{\bf Figure
2:}\quad (Top) Location of the image pair $a,b$ and the third
image $c$ in a pseudo-color image made from the {\em WFPC2}-F606W,
F814W and {\em ACS}-F850LP images. The red curves refers to the
critical lines of infinite magnification for sources placed at
$z$=5.576 and $z$=7.0 in the context of Kneib et al's (1996) mass
model revised to include the $z$=5.576 pair (shown as unlabelled
circles at the top of the figure) discussed by Ellis et al.\ (2001)
and a triply-imaged SCUBA source at $z$=2.515 (Kneib et al.\ 2004a).
(Bottom) Pseudo-color representation of the three images demonstrating
their association with a single lensed source.
\end{figure*}
\clearpage

\section{Redshift Constraints}

\subsection{Gravitational Lensing}

In the context of the tightly-constrained mass model of Abell 2218
(Kneib et al.\ 1996), updated to include the properties of the
$z$=5.576 pair identified by  Ellis et al.\ (2001) and the more
recent confirmation of a triply-imaged sub-millimeter selected
source at $z$=2.515 (Kneib et al.\ 2004a), the symmetry expected
for a lensed pair around the critical line implies a source
redshift $z_s>$6 (see curves in Figure~1). The absolute location
of the $z$=6 critical line is particularly well understood in this
region from the measured symmetry of the adjacent $z$=5.576 pair
around its critical line shown in Figure~2. However, the lensing
configuration for the new source provides only a fairly weak
constraint on the precise redshift beyond this lower limit since
the location of the critical line does not change significantly
beyond $z\simeq$6 (Figure~1).

Our mass model for Abell 2218 requires there to be a third image
of the source, which we successfully located in the $z'$-band
image at the expected position (image $c$ in Figure~2) and with
the expected flux (Table~1). Although our photometric coverage of
this third, fainter, image is not as complete as that for the
primary pair (because of the smaller field of the infrared cameras
used), importantly the {\em HST} photometry confirms the same
discontinuity in flux seen between F850LP and F606W (Figure~2,
Table~1). The improved mass model suggests that all 3 images
represent manifestations of a single source at $z>$6 magnified by
a factor of $\simeq$25 (in the case of images $a$ and $b$). The
intrinsic (unlensed) source brightness is $z_{850LP}=28.0\pm0.1$,
$H_{160W}=26.5\pm 0.1$.

\subsection{Spectral Energy distribution}

Figure~3 summarizes the available broad-band photometry for the
brightest of the three images ($a$ in Figure~1). We include, for
completeness, the Keck NIRC J measurement although, as a
ground-based measurement it is more adversely affected by crowding
and background issues related to the adjacent luminous cluster
members. Its low significance provides little more than evidence
for a detection. Accordingly, we do not use the J band data in any
of the subsequent analysis.

A significant discontinuity in flux is apparent in the wavelength
interval $\lambda\lambda$ $\sim$9200\AA--1$\mu$m. The overlap in
sensitivity between the WFPC-2 filter F814W and the {\em ACS}
filter F850LP is of particular diagnostic use. If it is assumed
the $z>$ 6 source has a UV continuum which rises to shorter
wavelengths with a discontinuity produced by a Gunn-Peterson
trough at around $\lambda_{rest}$=1216\AA, the ratio of the F814W
and F850LP fluxes can be used to estimate the redshift depending
on the slope of the UV continuum.

Assuming the spectral energy distribution is given by a simple
relation: $f(\lambda)\propto\lambda^{-\alpha}$ for
$\lambda/(1+z)>$1216\AA\ and $f(\lambda)=0$ otherwise, the
accurate {\em HST} photometry (F606W, F814W, F850LP, F160W)
implies the discontinuity occurs in the wavelength interval
9250--9850\AA, corresponding to 6.6$<z<$7.1 (see dashed lines in
Figure~3)\footnote{Although the weak NIRC J detection suggests a
shallower UV slope, we give this discrepancy low weight in view of
the superior quality of the HST data}. The photometric redshift
constraint is particularly firm at the lower end. Below
$z\simeq$6.6 it is hard to justify the observed F814W and F850LP
flux ratio regardless of the form of the spectral energy
distribution. In summary, therefore, the available HST photometry
suggests the lensed source lies beyond $z\simeq$6.6.

\begin{figure*}
\centerline{\psfig{file=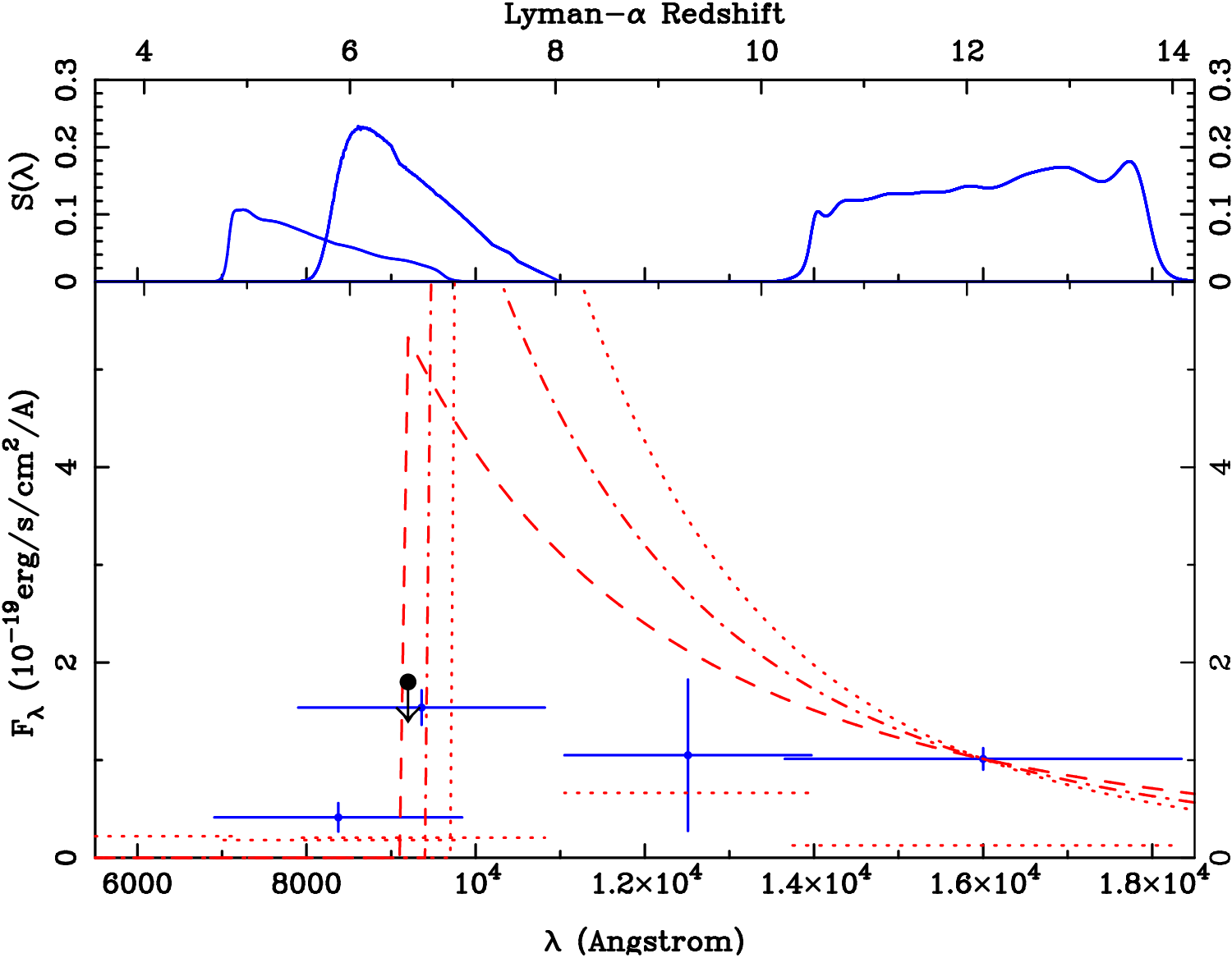,width=14cm,angle=-90}} \noindent{\bf
Figure 3:} (Top) Relative efficiencies of the filter+instrument
used in our photometric observations. From left to right: {\em
WFPC2}-F814W, {\em ACS}-F850LP and {\em NICMOS}-F160W. (Bottom)
Spectral energy distribution of image $a$ uncorrected for lensing
magnification. Photometric points are indicated by blue crosses,
and the 3$\sigma$ point source detection limits are indicated by a
horizontal dashed red lines below each data point. The
non-detection of the continuum in the LRIS 9000--9300\AA\ window
is indicated by the black arrow. Red dashed lines correspond to
power law spectral energy distributions (with
$f\propto\lambda^{-\alpha}$ for $\lambda/(1+z)>1216$\AA\ and $f=0$
otherwise) with slope indices (from bottom to top at 1.2 10$^4$
\AA\,) of $\alpha$=3,4 and 5. The available data are consistent
with a neutral hydrogen break in the interval 9250--9850\AA\
corresponding to 6.6$<z<$7.1.
\label{fig.sedplot}
\end{figure*}

\section{Optical and Infrared Spectroscopy}

Given the possibility that the lensed source lies beyond a
redshift $z\simeq$6.6 with a Gunn-Peterson discontinuity in the
wavelength range 9250--9850\AA\, we next tried to detect
Lyman~$\alpha$ emission in this region using both the Near
Infrared Spectrograph (NIRSPEC, McLean et al.\ 2001) on Keck II
and the Low Resolution Imaging Spectrograph (LRIS, Oke et al.\
1998) on Keck I. If around half of the flux in the F850LP band
arises via a Lyman~$\alpha$ emission line, in a manner analogous
to the $z=$5.576 source (Ellis et al. 2001), we considered that the
LRIS campaign should be successful in securing the redshift,
particularly since the lower part of the region to explore is,
by good fortune, one of the ``windows" of low OH sky emission used
by narrow band imagers to locate Lyman~$\alpha$
emitters (e.g. Kodaira et al.\ 2003).

We used LRIS on two runs, May 31 - June 1 2003 and June 30 - July
1 2003 (see Kneib et al.\ 2004b for a complete description of
these data). During the first observing run we observed the
brighter pair of the triple system using the 600 line grating
blazed at 1$\mu$m for a total of 13.8\ ksec. The first night
offered relatively poor conditions (cirrus, high humidity) but
9.0\ ksec was secured on the second night in slightly better
conditions. For these observations the wavelength coverage ranged
from 6930 to 9500\AA. No signal was detected from either image.

During the second observing run, we used the same configuration
and integrated for 9.2\ ksec in relatively good conditions with
$\sim 0.8\arcsec$ seeing. The wavelength coverage was extended
slightly to the red to reach 9600\AA. In the following, only the
second observing run will be utilized as the data quality is much
better. Flux calibration was conducted using Feige 67, Feige 110
and Wolf 1346 spectrophotometric standards observed in twilight.
The spectroscopic observations were reduced using {\sc IRAF}.
Despite the improved conditions, no continuum or emission lines
were detected to 9600\AA\, from either image. The absence of any
LRIS detection in the clean wavelength range 9000-9300 \AA\ places
an additional constraint on the lower redshift limit. In Figure 3
we illustrate this via an upper limit on the absolute flux (black
symbol) which further suggests the source lies beyond
$z\simeq$6.6.

The NIRSPEC (McLean et al.\ 1998) observations were obtained with
the Keck II telescope on the nights of May 10 and 11 2003.  The
spectrograph was used in low-dispersion mode using a filter with
transmission from 9500 to 11200\AA. A slit width of 0.76\arcsec\,
was used to acquire the pair giving a spectral resolution of
$R$=1100. We obtained 37 exposures of 900~sec yielding a total
integration time of 33\ ksec. We dithered along the slit between
two different positions, verifying our pointing on a nearby
reference star using the slit-viewing guide camera at each dither.
These observations were also reduced using {\sc IRAF}. Individual
spectra were registered using offsets determined from simultaneous
images obtained with the slit-viewing guide-camera. Our optimally
combined spectrum of each image comprised 26 individual exposures
totaling an effective exposure time of 23\ ksec. All of the
combined exposures were obtained in photometric conditions or
through thin clouds. The data were flux calibrated using
observations of Feige~110 (Massey \& Gronwall 1990) which provide
an effective calibration from 9700--10200\AA\ . Tests with two
other standard stars (Feige~34 and Wolf~1346) indicate the
relative flux calibration should be much more reliable than the
absolute calibration; we have probably suffered from slit losses
and possible absorption during the night due to faint cirrus. No
emission lines were seen in the extracted NIRSPEC spectra but a
faint stellar continuum was detected for both images $a$ and $b$;
the signal is slightly stronger for $a$.

Although observational selection plays an important role, as most
successfully identified $z>5$ galaxies display intense
Lyman~$\alpha$ emission (Spinrad 2003), it is interesting to
consider the maximum possible {\em observed} equivalent width,
$W_{max}$, implied by our non-detections in the LRIS and NIRSPEC
data. Using our spectrophotometric calibration, assuming a line
width of 5\AA\, and a conservative 5$\sigma$ detection limit,
Figure~4 shows $W_{max}$ as a function of wavelength for our data.
Although OH airglow emission precludes detection in a few small
wavelength regions, for 60\% of the important 9000\AA\,--9500\AA\,
interval an observed equivalent width larger than 120\AA\, (a
value much less than for most high redshift star forming sources,
Hu et al 2003) would have been detected. From 9000--9300\AA\,
(6.4$<z<$6.65) and longward of 9550\AA\, ($z>$6.85) the constraint
is tighter. Because of the much deeper exposures with NIRSPEC and
the stronger continuum redward of 9800\AA, a more stringent
5$\sigma$ upper limit of $W_{max}<$60\AA\, is derived for the
9550--11100\AA\, window effectively ruling out any reasonable
level of emission in the redshift range 6.85$<z<$8.2. Thus it
seems reasonable to deduce that, {\em if the redshift is not in
the 6.65$<z<$6.85 interval}, Lyman~$\alpha$ emission is either
very weak or absent. It will be important in future, deeper,
observations to rule out an emission line hiding in the OH forest
at $\sim$9500\AA\, corresponding to a source at $z\sim$6.8.

\begin{figure*}
\centerline{\psfig{file=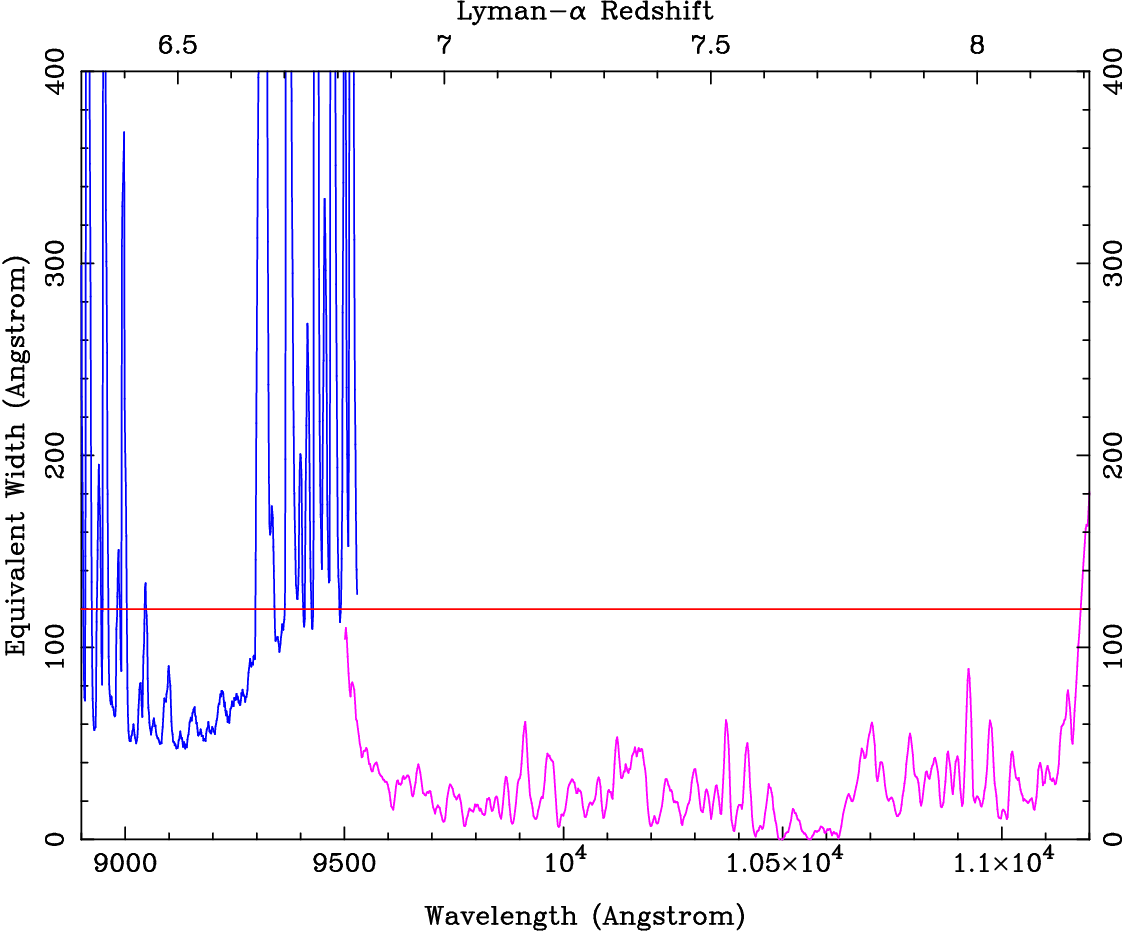,width=15cm,angle=270}}

\noindent{\bf Figure 4} Observed equivalent width detection limit
(5$\sigma$), $W_{max}$, for Lyman~$\alpha$ emission in the LRIS
and NIRSPEC data assuming a line width of 5\AA\,. An emission line
stronger than $W$=120\AA (indicated by the horizontal line),
corresponding to an integrated line flux of 1.6$\times$10$^{-18}$
ergs s$^{-1}$ cm$^{-2}$\AA\,$^{-1}$, would have been seen for 60\%
of the wavelength range 9000\AA\,--9500\AA. Constraints in the
range 9000--9300\AA\, and beyond 9550\AA\ are tighter. The longer
NIRSPEC integrations provide a maximum equivalent width below
$W$=60\AA\, for the entire 9550--11100\AA\, range, corresponding
to an integrated line flux of 7$\times$10$^{-19}$ ergs s$^{-1}$
cm$^{-2}$\AA\, $^{-1}$. \label{fig.eqw}
\end{figure*}

The absence of Lyman~$\alpha$ emission in a distant source may
seem surprising. However, there are many examples of luminous
Lyman break galaxies at lower redshift with weak or no emission
(Shapley et al.\ 2003). Indeed, those authors claim only 20-25\%
of star-forming examples show Lyman~$\alpha$ emission sufficiently
prominent to be classified as narrow band excess objects.

The most important outcome of the long NIRSPEC exposure is the
detection of a very faint continuum redward of 9800\AA\,
(Figure~5). A continuum signal is seen in both images. As it is
stronger in image $a$, we will use this spectrum for the following
analysis. Given the absence of Lyman~$\alpha$ emission and the
inference of a photometric break in the wavelength range
9250--9850\AA\, the key question is therefore the lowest
wavelength at which a continuum signal can be seen in the NIRSPEC
data.

Inspection of the data reveals a drop in the continuum flux at
9800\AA\, shortward of which there is no reliably detected signal
despite a robust photometric calibration (see upper panels of
Figure 5). The OH spectrum is fairly clean in this region although
there is some atmospheric absorption which could affect the
calibration at 9500--9700\AA\ . Although the feature itself is
only marginal, the absence of flux below 9800\AA\ seems significant
when one considers the photometric data summarised in Figure 3.
Specifically, if the UV continuum extended down to 9250\AA\,, as
would be the case if the source were at $z$=6.6, it is difficult
to understand why the stellar continuum is not detected to shorter
wavelengths (Figure 5). Although this weak feature is the only
indicator in our exhaustive attempts to measure a spectroscopic
redshift, if it is indeed the cause of the Gunn-Peterson edge
inferred from the photometric data, a redshift of $z\simeq$7.05 is
implied. In this case, the photometric redshift analysis discussed
in $S\S$3.2 would indicate a steep UV continuum slope of
$\alpha$=5.

\clearpage
\begin{figure*}
\centerline{\psfig{file=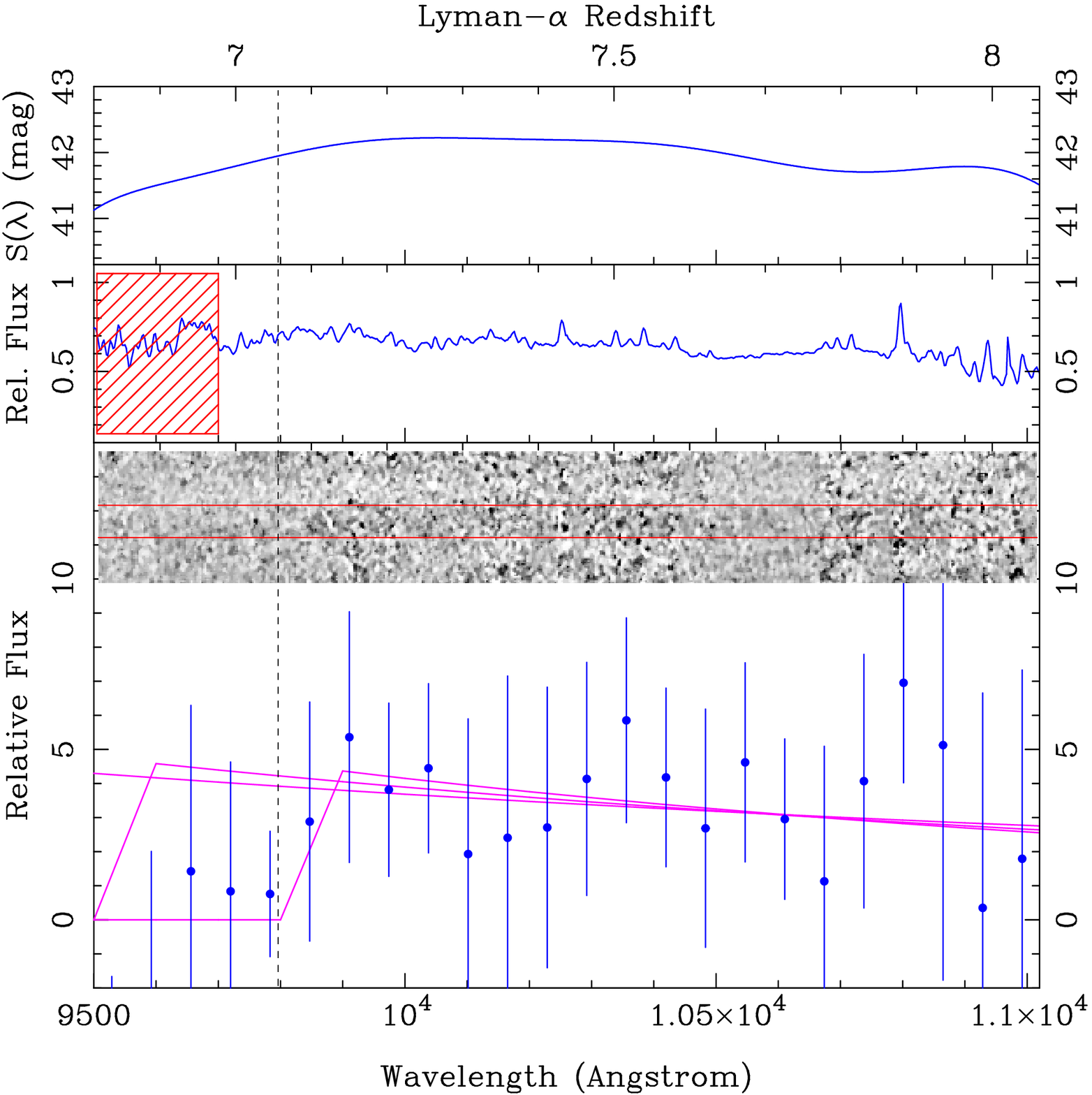,width=18cm,angle=0}} \noindent{\bf
Figure 5} NIRSPEC spectrum for image $a$ in the wavelength region
9500\AA\,--1.1$\mu$m. (Top) Relative sensitivity of NIRSPEC in the
configuration used derived from 3 independent flux calibration
stars. Adequate sensitivity is available over the entire range.
(Middle) {\rm observed} spectrum of one of these standards,
Feige~110 illustrating {\sl a priori} the absence of any strong
atmospheric features in the reduction. The red hashed rectangle
indicates the region where variable atmospheric absorption is
expected (Massey \& Gronwall 1990). (Bottom) Flux calibrated
binned spectrum for image $a$. Error bars include the effect of
wavelength-dependent OH emission. Curves represent UV SEDs
consistent with the photometry (Figure 3) spanning the range
6.6$<z<$7.1. The absence of significant flux below 9800 \AA\
suggests a redshift at the upper end of this range, i.e.
$z\simeq$7.

\label{}
\end{figure*}
\clearpage
\section{A z$\simeq$7 Source}

Clearly identifying the redshift of this source has been extremely
challenging and further studies would be highly desirable to
confirm our conclusions. We have introduced three independent
arguments which, taken in combination, justify why the
newly-located multiple images arise via a single lensed source at
a redshift $z\sim$7. First, the geometrical configuration of
images $a,b$ and $c$ with respect to the critical line in this
well-modeled cluster indicates $z>$6. We argued that the location
of the critical line in the vicinity of the pair is tightly
constrained from the successful identification of the earlier
source at $z$=5.576 (Ellis et al 2001). Secondly, there is a strong
break indicated in the spectral energy distribution (SED)
delineated by the broad-band photometry extending from the F606W
to F160W ($H$) filters, arguing for a redshift $6.6<z<7.1$.
The lower redshift limit $z>$6.6 is particularly firm and
supported by the absence of any LRIS continuum signal in the
wavelength range 9000-9300 \AA. Finally the absence of any
detectable signal below 9800\AA\ seen in the NIRSPEC continuum
argues the redshift lies in the upper end of the photometric
range, $z\simeq$7, as otherwise the rising UV continuum would have
been detected.

Even if one disregards the weak feature in the NIRSPEC spectrum,
the 6.6$<z<$ 7.1 source is of considerable interest since, as we
discussed in $\S$3-4, it is difficult to reconcile a spectral
break at $\lambda>$9250\AA\ with a UV continuum slope of
$\alpha\lapprox 3$, assuming the UV continuum is described by a power law
above the Lyman~$\alpha$ discontinuity (Figure 3). Starburst
models assuming Population II metallicity and a Salpeter stellar
initial mass function (IMF) typically produce slopes of
$\alpha\lapprox 2$ (Leitherer et al.\ 1999). Clearly the inferred
UV continuum of the $z\sim7$ source rises faster to shorter
wavelengths than for any of the models calibrated with local data.
Models based on massive metal-free stars, such as those which may
be expected close to the epoch of reionization (Abel, Bryan, \&
Norman 2000; Bromm, Coppi, \& Larson 1999), do produce somewhat
steeper UV slopes, $\alpha\simeq3$ (Schaerer 2002). More precise
photometry for this object, particularly in the 1--1.5$\mu$m range 
is needed to understand this issue. If a steep UV slope ($\alpha\geq 3$) is
indeed confirmed, the source may represent a promising candidate
for a Population III starburst.

The star-formation rate (SFR) of the source can be estimated from
the stellar UV continuum luminosity.  The unlensed F160W
magnitude, $H_{160W}=26.5\pm0.1$, translates into a specific flux
of $2.35\times10^{-31}$erg~s$^{-1}$cm$^{-2}$Hz$^{-1}$ at
1.6$\mu$m, assuming a smooth SED.  The intrinsic specific
luminosity at 2000\AA\ is then $1.85\times10^{28}$erg\,s$^{-1}$
Hz$^{-1}$ implying a SFR of 2.6~M$_\odot$yr$^{-1}$ using
Kennicutt's (1998) empirical calibration. We estimate a 15\% error
arising from uncertainties in the photometry and magnification.
This derived SFR is significantly higher than that for the
$\simeq$0.5 $M_{\odot}$ yr$^{-1}$ computed for the lensed source
at $z$=5.576 (Ellis et al.\ 2001) but within the range of typical
Lyman-$\alpha$ emitters at $z\simeq$5.5-6.5.

Our estimated SFR may suffer from two systematic errors.  If dust
is present, it will dim the UV continuum, causing us to
underestimate the value.  However, assuming a familiar selective
extinction curve, reddening would then imply the intrinsic UV
continuum slope is even steeper than $\alpha\simeq$3-5. A second
uncertainty arises from our assumed stellar initial mass function
(IMF) and metallicity via Kennicutt's local calibration. If the
IMF is top-heavy or the metallicity is lower than in local
starbursts, the SFR will likely have been over-estimated.

Given the intense star formation, the apparent lack of
Lyman~$\alpha$ emission is puzzling if dust extinction is not
important. Weak or absent Lyman~$\alpha$ emission seen in the
younger ($<$1 Gyr) Lyman break galaxies at $z\simeq$3 has been
interpreted via the presence of dust shrouds which are eventually
disrupted via feedback processes as the stellar population matures
(Shapley et al. 2003). As our source is likely younger than 500
Myr, a very specific dust/gas geometry would be needed to strongly
extinct the Lyman~$\alpha$ photons without reddening the UV
continuum. This problem may be exacerbated if the IMF is top-heavy
or the metallicity low, since the ionizing photon production rate
will be higher making any emission line yet more prominent (c.f.
Malhotra \& Rhoads 2002).

An alternative explanation for the absence of Lyman $\alpha$
emission may be incomplete reionization at $z\simeq$7. Neutral
hydrogen can scatter Lyman~$\alpha$ photons after they have
escaped the ISM of the emitting galaxy (e.g., Miralda-Escude \&
Rees 1998). If the source were embedded in a neutral zone, even a
strong Lyman~$\alpha$ emission line emitted from the galaxy could
be quenched. However, in such cases, simulations suggest that
emission can still be observed even from sources embedded in fully
neutral zones, depending on details of the many relevant
parameters of the source (Haiman 2002, Santos 2003).

Since the universe is only ~750 Myr old at $z=7$, the total
stellar mass is unlikely to exceed $1-2\times10^9$~M$_\odot$.
Halos of mass $>1.2\times10^{10}$M$_\odot$ at $z\simeq$7 are quite
consistent with conventional structure formation models (Barkana
\& Loeb 2002) and sufficient to supply baryons to sustain the
observed star formation rate. The instantaneous baryonic accretion
rate for such a halo at $z=7$ is, on average,
4.4~M$_\odot$yr$^{-1}$ (e.g. Lacey \& Cole 1993), so even if an
earlier starburst consumed most of the baryons, they can be
rapidly replenished. Although we have no constraints on the actual
star-formation history, these arguments emphasize that we are not
necessarily viewing the system at a special time in its evolution.

The strong magnification ($\simeq\times25$) of the two brighter
images gives us our first glimpse into the morphological structure
of a very distant source on sub-kpc scales. Both images $a$ and
$b$ have a bright core, a second fainter knot, and extended
emission of lower surface brightness (Figure~1). There is no
noticeable color gradient. Our mass model for Abell 2218 implies
magnifications of $\sim\times15$ along the major axis of images
$a$ and $b$, and $\sim\times1.7$ along the minor axis. The
$0.15\arcsec$ width of the F850LP PSF translates into a physical
resolution of $\sim$470~pc along the minor axis and $\sim$50~pc
along the major axis.

We thus infer that the source (observed with 3.6\arcsec$\times
<0.15$\arcsec) has a maximum physical size of 1.2~kpc by
$<$500~pc. The maximal associated area of 0.6~kpc$^2$ indicates a
star-formation surface density in excess of
4.3~M$_\odot$yr$^{-1}$kpc$^{-2}$. The bright knot is only 100~pc
across implying a star-formation surface density in the range
$\sim$50-250~M$_\odot$yr$^{-1}$kpc$^{-2}$ depending on the
exact geometry, comparable to the most intense starburst activity
observed locally (Kennicutt 1998).

If this source is typical of those which reionized the Universe in
a narrow time interval of $\Delta\,z$=1 around $z\simeq$7, we can
estimate the expected surface density from the arguments presented
by Stiavelli et al (2003). Depending on the source temperature,
Lyman continuum escape fraction and clumpiness of the IGM, we
would deduce surface densities $n\simeq$0.3-5 arcmin$^{-2}$ are
necessary. In as much as it is possible to estimate the actual
surface density from one source and the limited area examined by
looking through only one cluster lens to find it, we find a number
density of $n\simeq1\pm0.5$ arcmin$^{-2}$.

In conclusion, we present evidence that we have found a highly
magnified source which lies beyond $z\simeq6.6$, possibly at
$z\simeq$7.  Even in advance of the infrared capabilities of
\textit{JWST}, further observation of this source will be
important in determining a more secure redshift, and improving the
constraints on the slope of the UV continuum.  Further
spectroscopy in the 9200\AA -1$\mu$m region would be valuable to
probe any Lyman~$\alpha$ emission at $z\simeq$6.8 (Figure 4) and
to confirm (or otherwise) the significance of the weak continuum 
drop seen in the NIRSPEC data at 9800\AA\ . The location of the 
Gunn-Peterson trough might also be verified more precisely via narrow band
imaging through gaps in OH forest and with a deep ACS grism spectroscopy.

Notwithstanding the need for further work and regardless of its
precise redshift in the constrained window 6.6$<z<$7.1, our source
appears to be a star-forming galaxy with intriguing spectral
properties, possibly representative of a new population
responsible for ending cosmic reionization. While our discovery
highlights many of the challenges facing searches for those
$z>$6.5 galaxies responsible for reionization, it also
demonstrates the ability of strong lensing by clusters of galaxies
to locate and reveal the detailed properties of high redshift
sources. The lensed galaxy presented in this paper, if observed
unlensed at $z_{AB}\sim28.5$ would lie at the detection limit of
the upcoming UDF observation\footnote{\rm
http://www.stsci.edu/hst/udf/parameters} and no spectroscopic
follow-up would have been possible.

\acknowledgments

We thank Fred Chaffee for his continued encouragement to track
down the nature of this intriguing system and acknowledge useful
discussions with Bob Abraham, Chris Conselice, Graham Smith, Mark
Sullivan and Tommaso Treu. Hy Spinrad and Andy Bunker are thanked
for reading an earlier version of this manuscript and offering
valuable suggestions. We thank two anonymous referees for their
valuable comments which significantly improved the presentation of
our data. Alice Shapley and Dawn Erb for helpful advice on the
optimum procedures for obtaining and reducing faint NIRSPEC data.
We also thank James Larkin and James Graham for their advice
concerning the use of NIRSPEC in the 1$\mu$m wavelength region.
Andrew Blain and Naveen Reddy kindly provided access to the Keck
NIRC observations. Faint object spectroscopy at the Keck
observatory is made possible through the dedicated efforts of Ian
McLean and collaborators for NIRSPEC, and Judy Cohen, Bev Oke,
Chuck Steidel and colleagues at Caltech for LRIS. JPK acknowledges
support from Caltech and CNRS. MRS acknowledges the support of
NASA GSRP grant NGT5-50339. The study of Abell 2218 as a cosmic
lens is supported by NASA STScI grant HST-GO-09452.01-A.

\newpage
\begin{deluxetable}{rlll}
\tabletypesize{\normalsize} \tablecaption{Observed Photometry for
the Triple System}\tablewidth{0pt} \tablehead{ \colhead{}  &
\colhead{a} & \colhead{b} & \colhead{c} } \startdata
$\alpha_{J2000}$ & 16:35:54.73 & 16:35:54.40 & 16:35:48.92 \\
$\delta_{J2000}$ & 66:12:39.00 & 66:12:32.80 & 66:12:02.45 \\
$V_{606W}$  & not detected & not detected & not detected \\
$I_{814W}$  & 26.5$\pm0.2$   & 26.4$\pm0.2$   & not detected \\
$z_{850LP}$ & 24.38$\pm0.05$ & 24.54$\pm0.05$ & 25.9$\pm0.1$ \\
$J$         & 23.4$\pm0.3$   & 23.5$\pm0.3$  & --- \\
$H_{160W}$  & 22.96$\pm0.06$ & 23.01$\pm0.07$ & --- \\
$V_{606W}$-$z_{850LP}$ & $>$3.6 & $>$3.5 & $>$2.1 \\
$I_{814W}$-$z_{850LP}$ & 2.1$\pm0.2$ & 1.9$\pm0.2$ & $>$1.8\\
$z_{850LP}$-$J$ & 1.0$\pm0.3$ & 1.0$\pm0.4$ & --- \\
$z_{850LP}$-$H_{160W}$ & 1.42$\pm0.1$ & 1.53$\pm0.1$ & --- \\
Magnification & 25$\pm 3$  & 25$\pm 3$  & 5.3$\pm 0.5$ \\
\enddata

\bigskip

\leftline{$3\sigma$ detection limits for a point source:}

\smallskip

\leftline{$V_{606W}=28.0$, $I_{814W}=27.2$, $z_{850LP}=26.7$,
$J=24.4$, $H_{160W}=25.8$.}

\end{deluxetable}


\begin{thebibliography}{}

\bibitem[Abel, Bryan, \& Norman(2000)]{2000ApJ...540...39A} Abel, T.,
Bryan, G.~L., \& Norman, M.~L.\ 2000, {\it Astrophys. J.}, {\bf 540}, 39
\bibitem[]{664} Barkana, R., Loeb, A., 2002, {\it Astrophys. J.}, {\bf 578}, 1.
\bibitem[]{665} Becker, R. H. et al.\ (SDSS Collaboration) 2001, {\it Astron. J.}, {\bf 122}, 2850.
\bibitem[]{666} Bouwens, R. J. et al, 2001, {\it Astrophys. J.}, {\bf 595}, 589.
\bibitem[Bromm, Coppi, \& Larson(1999)]{1999ApJ...527L...5B} Bromm, V.,
Coppi, P.~S., \& Larson, R.~B.\ 1999, {\it Astrophys. J. Lett}, {\bf 527}, L5
\bibitem[]{669} Dickinson, M. et al, 2003, {\it Astrophys. J. Lett} sumitted (astro-ph/0309070)
\bibitem[]{670} Djorgovski, S.G., Castro, S.M., Stern, D. \& Mahabal,
A.A. 2001, {\it Astrophys. J. Lett}, {\bf 560}, L5.
\bibitem[]{672} Ebbels, T., Ellis, R.S., Kneib, J-P., Leborgne, J-F.,
Pell\`o, R., Smail, I.R. \& Sanahuja, B. 1999, {\it Mon. Not. R.
astr. Soc.}, {\bf 295}, 75.
\bibitem[]{675} Ellis, R.S., Santos, M.R.,  Kneib, J-P, Kuijken, K., 2001,
{\it Astrophys. J. Lett}, {\bf 560}, L119.
\bibitem[]{677} Fan, X. et al.\ (SDSS Collaboration) 2002, {\it Astron.
J.}, {\bf 123}, 1247.
\bibitem[Haiman(2002)]{2002ApJ...576L...1H} Haiman, Z.\ 2002, {\it Astrophys. J. Lett},
{\bf 576}, L1
\bibitem[]{681} Hu, E.M., Cowie, L.L. \& McMahon, R.G. 1998, {\it
Astrophys. J.}, {\bf 502}, L99.
\bibitem[]{683} Hu, E.M., Cowie, L.L., Capak, P., McMahon, R.G., Hayashimo, T.,
Komiyama Y., 2004, to appear in {\it Astron. J.}
(astro-ph/0311528)
\bibitem[]{686} Kennicutt, R.C. 1998, {\it Ann. Rev. Astron. Astr.},
{\bf 36}, 189.
\bibitem[]{688} Kneib, J.-P, Ellis, R.S., Smail, I.R., Couch, W.J. \&
Sharples, R.M. 1996, {\it Astrophys. J.}, {\bf 471}, 643.
\bibitem[]{690} Kneib, J.-P., van der Werf, P., Kraiberg, K., Smail, I.,
Blain, A., Frayer, D., Barnard, V., Ivison, R., 2004a, {\it Mon. Not. R.
astr. Soc.}, in press.
\bibitem[]{693} Kneib, J.-P. et al, 2004b, in preparation.
\bibitem[]{694} Kodaira, K. et al.\ 2003, {\it P.A.S.J. Lett}, {\bf 55}, L17.
\bibitem[]{695} Kogut, A. et al.\ 2003, {\it Astrophys. J. Suppl.}, {\bf 148}, 161.
\bibitem[Lacey \& Cole(1993)]{1993MNRAS.262..627L} Lacey, C.~\& Cole, S.\
1993, {\it Mon. Not. R. astr. Soc.}, {\bf 262}, 627
\bibitem[]{698}Leitherer, C., Schaerer, D., Goldader, J.D. 1999,
{\it Astrophys. J. Suppl.}, {\bf 123}, 3.
\bibitem[]{700} Malhotra, S. et al.\ 2001, in {\it Gas and Galaxy
Evolution}, eds. Hibbard, J.E. et al, ASP Conf. Series
(astro-ph/0102140)
\bibitem[Malhotra \& Rhoads(2002)]{2002ApJ...565L..71M} Malhotra, S.~\&
Rhoads, J.~E.\ 2002, {\it Astrophys. J. Lett}, {\bf 565}, L71
\bibitem[Massey \& Gronwall(1990)]{mas90} Massey, P.~\& Gronwall, C.\
1990, {\it Astrophys. J.}, {\bf 358}, 344
\bibitem[McLean et al.(1998)]{mcl98} McLean, I.~S.~et al.\ 1998,
\procspie, {\bf 3354}, 566
\bibitem[Miralda-Escude \& Rees(1998)]{Miralda-Escude1998}
        Miralda-Escude, J.~\& Rees, M.~J.\ 1998,{\it Astrophys. J.}, {\bf 497}, 21
\bibitem[]{711} Oke, J.B. et al.\ 1995, {\it Publ. Astr. Soc. Pac.},
{\bf 107}, 375.
\bibitem[]{713} Santos, M.R., 2003, {\it Mon. Not. R. astr. Soc.}, sumitted astro-ph/0308196
\bibitem[]{714} Santos, M.R., Ellis, R.S., Kneib, J-P, Richard, J., Kuijken, K., 2003,
{\it Astrophys. J.} sumitted (astro-ph/0310478)
\bibitem[]{716} Schaerer, D., 2002, {\it Astronomy \& Astrophysics}, {\bf 382}, 28.
\bibitem[]{718} Shapley, A.E., Steidel, C.C., Pettini, M., Adelberger, K.L., 2003, {\it
Astrophys. J.}, {\bf 588}, 65.
\bibitem[]{720} Spergel, D.N. et al.\ 2003, {\it Astrophys. J. Suppl.}, {\bf 148}, 175.
\bibitem[]{721} Spinrad, H., 2003, in {\it Astrophysics Update}, Mason J. (Ed.)
(astro-ph/0308411)
\bibitem[]{723} Spinrad, H., Stern, D., Bunker, A., Dey, A.,
Lanzetta, K., Yahil, A., Pascarelle, S., Fernandez-Soto, A. 2003,
{\it Astron. J}, {\bf 116}, 2617.
\bibitem[]{726} Stanway, E., Bunker, A., Mc Mahon, R., Ellis, R.S., Treu, T., Mc Carthy, P., 2003,
{\it Astrophys. J.} sumitted (astro-ph/0308124)
\bibitem[]{728} Stern, D. \& Spinrad, H. 1999, {\it Publ. Astron. Soc.
Pac.}, {\bf 111}, 1475.
\bibitem[]{730} Stiavelli, M., Fall, S.M., Panagia, N.\ 2004, {\it
Astrophys. J.}, in press.
\bibitem[Weymann et al.(1998)]{1998ApJ...505L..95W} Weymann, R.~J., Stern,
D., Bunker, A., Spinrad, H., Chaffee, F.~H., Thompson, R.~I., \&
Storrie-Lombardi, L.~J.\ 1998, {\it Astrophys. J. Lett}, {\bf 505}, L95
\bibitem[]{735} Yan, H., Windhorst, R.A., Cohen, S.H., 2003,
{\it Astrophys. J. Lett}, {\bf 585}, L93.

\end{thebibliography}
\end{document}